\documentclass[sort&compress]{aipproc}

\layoutstyle{6x9}
\usepackage{epsfig}

\def\p1{\phantom{1}}

\def\simless{\mathbin{\lower 3pt\hbox
     {$\rlap{\raise 5pt\hbox{$\char'074$}}\mathchar"7218$}}}   
\def\simmore{\mathbin{\lower 3pt\hbox
     {$\rlap{\raise 5pt\hbox{$\char'076$}}\mathchar"7218$}}}   
\def\hide#1{}
\newcommand{\msun}{\ensuremath{{\rm M}_\odot}}

\begin{document}

\title{The elusive Innermost Stable Circular Orbit: Now you see it, now
you don't}


\author
{Mariano M\'endez}
{address={SRON - Netherlands Institute for Space Research, Sorbonnelaan 2, 3584 
CA Utrecht, The Netherlands},
altaddress={Astronomical Institute, University of Amsterdam, Kruislaan
403, 1098 SJ Amsterdam, The Netherlands}
}

\begin{abstract} I study the behaviour of the maximum rms fractional
amplitude, $r_{\rm max}$ and the maximum coherence, $Q_{\rm max}$, of
the kilohertz quasi-periodic oscillations (kHz QPOs) in a dozen low-mass
X-ray binaries. I find that: (i) The maximum rms amplitudes of the lower
and the upper kHz QPO, $r^{\ell}_{\rm max}$ and $r^{\rm u}_{\rm max}$,
respectively, decrease more or less exponentially with increasing
luminosity of the source; (ii) the maximum coherence of the lower kHz
QPO, $Q^{\ell}_{\rm max}$, first increases and then decreases
exponentially with luminosity; (iii) the maximum coherence of the upper
kHz QPO, $Q^{\rm u}_{\rm max}$, is more or less independent of
luminosity; and (iv) $r_{\rm max}$ and $Q_{\rm max}$ show the opposite
behaviour with hardness of the source, consistent with the fact that
there is a general anticorrelation between luminosity and spectral
hardness in these sources. Both $r_{\rm max}$ and $Q_{\rm max}$ in the
sample of sources, and the rms amplitude and coherence of the kHz QPOs
in individual sources show a similar behaviour with hardness. This
similarity argues against the interpretation that the drop of coherence
and rms amplitude of the lower kHz QPO at high QPO frequencies in
individual sources is a signature of the innermost stable circular orbit
around a neutron star. \end{abstract}

\maketitle


\section{Introduction}
\label{intro}

Kilohertz quasi-periodic oscillations (kHz QPOs) in the X-ray flux of
low-mass X-ray binaries have drawn much attention since their discovery,
about ten years ago. The reason for this continued interest is that
since they most likely reflect the motion of matter very close to the
neutron star (or black hole), these QPOs may provide one of the few
direct ways of measuring effects that are unique to the strong
gravitational-field regime in these systems. Often two simultaneous kHz
QPOs are seen in the power density spectra of low-mass X-ray binaries,
the lower and the upper kHz QPO according to how they appear sorted in
frequency.

Most of the work on QPOs in these years \citep[see][for a
review]{vanderklis-review} has focused on the frequencies of these QPOs,
because those frequencies provide insights into the dynamics of the
system. For instance, different from the Newtonian theory of
gravitation, in general relativity the effective potential as a function
of radial distance to the central source has a maximum. A particle in a
circular orbit at that radius would be in unstable equilibrium; if
perturbed, the particle would fall onto the central object. This radius
defines the innermost stable circular orbit, or ISCO. No stable orbit
around the central object is possible inside the radius of the ISCO,
which in the Schwarzschild case (non-rotating central object) is $r_{\rm
ISCO} = 6 G M/c^2$ \citep*{bardeen}. If (at least) one of the QPOs
reflects Keplerian motion around the neutron star \citep*{miller,
stella2}, the ISCO would set a limit to the maximum frequency that the
QPOs can reach \citep{miller}. By measuring that maximum frequency it
would be possible to infer the mass and constrain the radius of the
neutron star.

From the beginning there has been interest on the other two properties
of the kHz QPOs, their amplitude and coherence
\citep[e.g.,][]{vanderklis-scox-1}, but systematic studies of those
other QPO properties only started to take off slightly later
\citep*[e.g.][]{jonker-340+0, disalvo-1728, vanstraaten-0614,
mendez-3srcs}.

Recently, Barret et al. \cite{barret-1608, barret-1636, barret-nordita}
carried out a systematic study of the kHz QPO coherence and rms
amplitude in three X-ray binaries, 4U 1636--53, 4U 1608--52, and 4U
1735--44. They find that in all three sources the coherence and rms
amplitude of the lower kHz QPO increase slowly with frequency, and after
the coherence and rms amplitude reach their maximum values, they
decrease abruptly as the QPO frequency keeps on increasing. (The sudden
drop is most noticeable in the coherence of the lower kHz QPO; in the
case of the rms amplitude the drop is less abrupt.) Barret et al.
propose that this behaviour is due to effects related to the ISCO around
the neutron star in these systems.

Triggered by these results, I investigated the dependence of the maximum
coherence and rms amplitude of both kHz QPOs in a large sample of
low-mass X-ray binaries.  


\section{Data selection and analysis}
\label{data}

All the data that I use here were obtained over the last 10 years with
the Proportional Counting Array (PCA) on board the Rossi X-ray Timing
Explorer (RXTE). I collected most of these data from the literature. For
this I searched, for as many sources as possible, all published values
of the rms fractional amplitude and the coherence of the kHz QPOs.
\citep[see][for a detailed description of the selection and
characteristics of the data.]{mendez-06} The coherence $Q$ of a QPO is
defined as $Q = \nu_{\rm QPO} / \lambda$, where $\nu_{\rm QPO}$ and
$\lambda$ are the frequency and the full-width at half-maximum of the
QPO. The rms amplitude, $r$, is calculated from $P$, the integral from
$0$ to $\infty$ of the Fourier power under the QPO, and the source
intensity, $S$, as $r = 100 \times \sqrt{P/S}$; from this definition,
$r$ is expressed as a percent of the total intensity of the source.


Here I use the naming convention of \cite*{belloni-2002}, in which the
lower kHz QPO is called $L_{\ell}$, and the upper kHz QPO is called
$L_{\rm u}$. The frequency, coherence and amplitude of each QPO carry a
subscript or superscript $\ell$ or $u$, respectively. I use the
expression ``kilohertz QPO'' to refer to features in the power spectra
of neutron star systems that have frequencies $> 150$ Hz, and that have
not been identified as hectohertz QPO \cite[e.g.,][]{vanstraaten-0614}.

Once I have collected all $Q_{\rm max}$ and $r_{\rm max}$ values for
each QPO for each source, and the frequencies $\nu_{\ell}$ and $\nu_{\rm
u}$ at which those maximum values occur, I use Figure 1 in
\cite{ford-et-al-2000} to get the corresponding source luminosity: Using
the QPO frequency as input, I read off the luminosity of the source at
that frequency from that Figure. For three sources, 4U 1608--52, 4U
1636--53, and 4U 1820--30, the maximum rms amplitudes of $L_{\rm u}$
occur in states of the source in which the QPO frequencies are outside
the range of frequencies in Figure 1 of \cite{ford-et-al-2000}. In those
three cases I either search the literature for flux measurements of the
source in that state, or I extract spectra from the corresponding {\em
RXTE} observations, and calculate the luminosities myself, or both. 


The luminosities used here are uncertain for the following reasons: (i)
Statistical error in the fluxes derived from model fitting. These errors
are usually less than $5 - 10\%$. (ii) Accuracy in the determination of
the luminosity for a given QPO frequency using the results of
\cite{ford-et-al-2000}. The reason for this is the so-called
parallel-track phenomenon \cite[e.g.,][]{mendez-1608}, which introduces
an uncertainty of the order of $20 - 50\%$ in the luminosity
\citep{mendez-3srcs}. (iii) The use of the $2-50$ keV flux as a measure
of the bolometric flux of the source. This effect contributes
uncertainties of the order of $20-25\%$ \citep[see][]{ford-et-al-2000}.
(iv) Uncertainty in the distance to these sources \citep[see][for a
discussion in the context of these sources]{ford-et-al-2000}; this can
yield uncertainties of up to 60\% in the luminosity \citep[see,
e.g.,][]{christian-swank}. In this paper I use a fixed error of $25\%$
in the values of $L/L_{\rm Edd}$ as indicative of the error in the
luminosity. It is clear from the previous discussion that this is a
lower limit to the real error in this quantity.


\begin{figure*}
\centerline{\epsfig{file=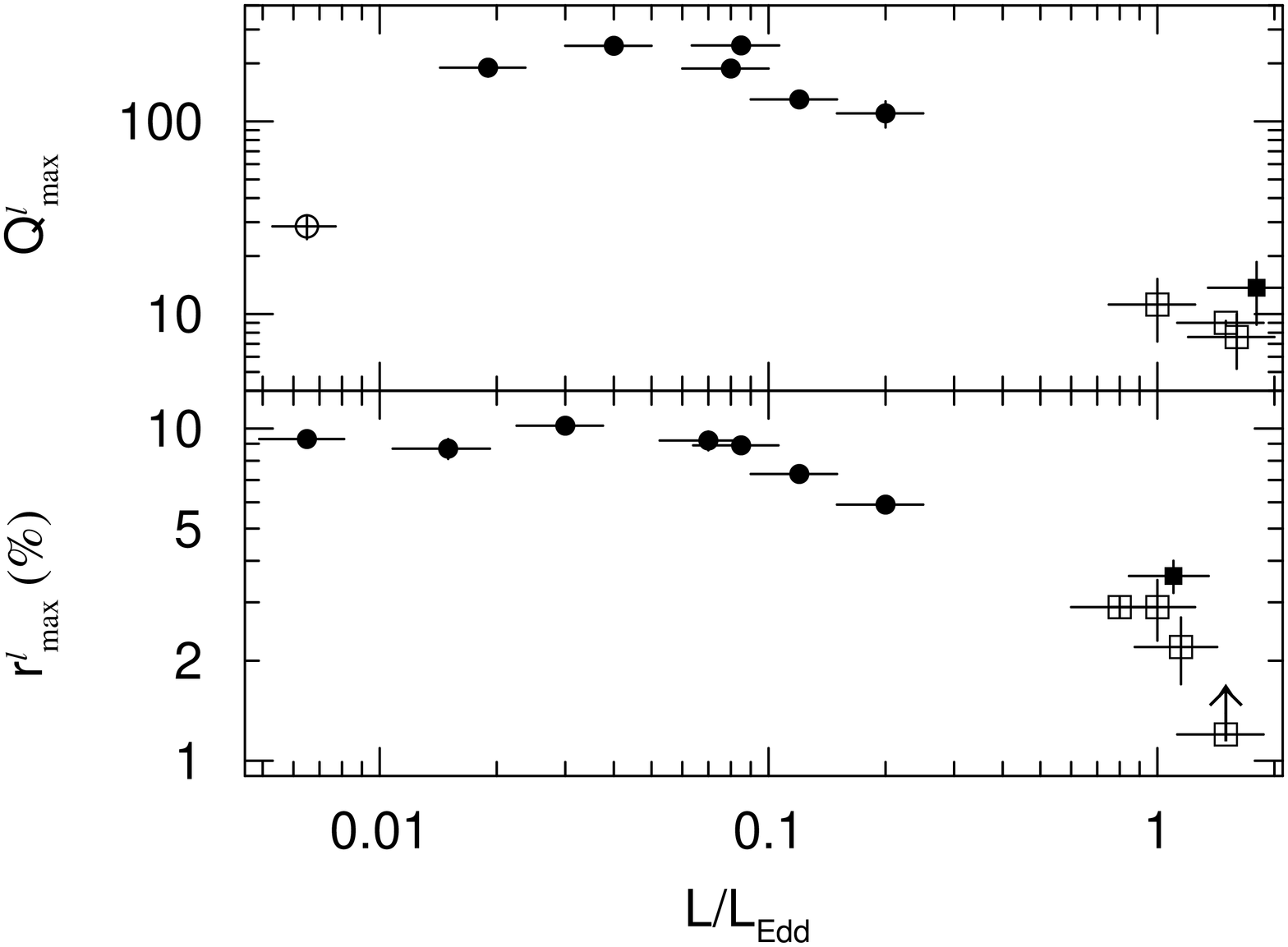,angle=-90,width=7.0cm}~~~~~~
            \epsfig{file=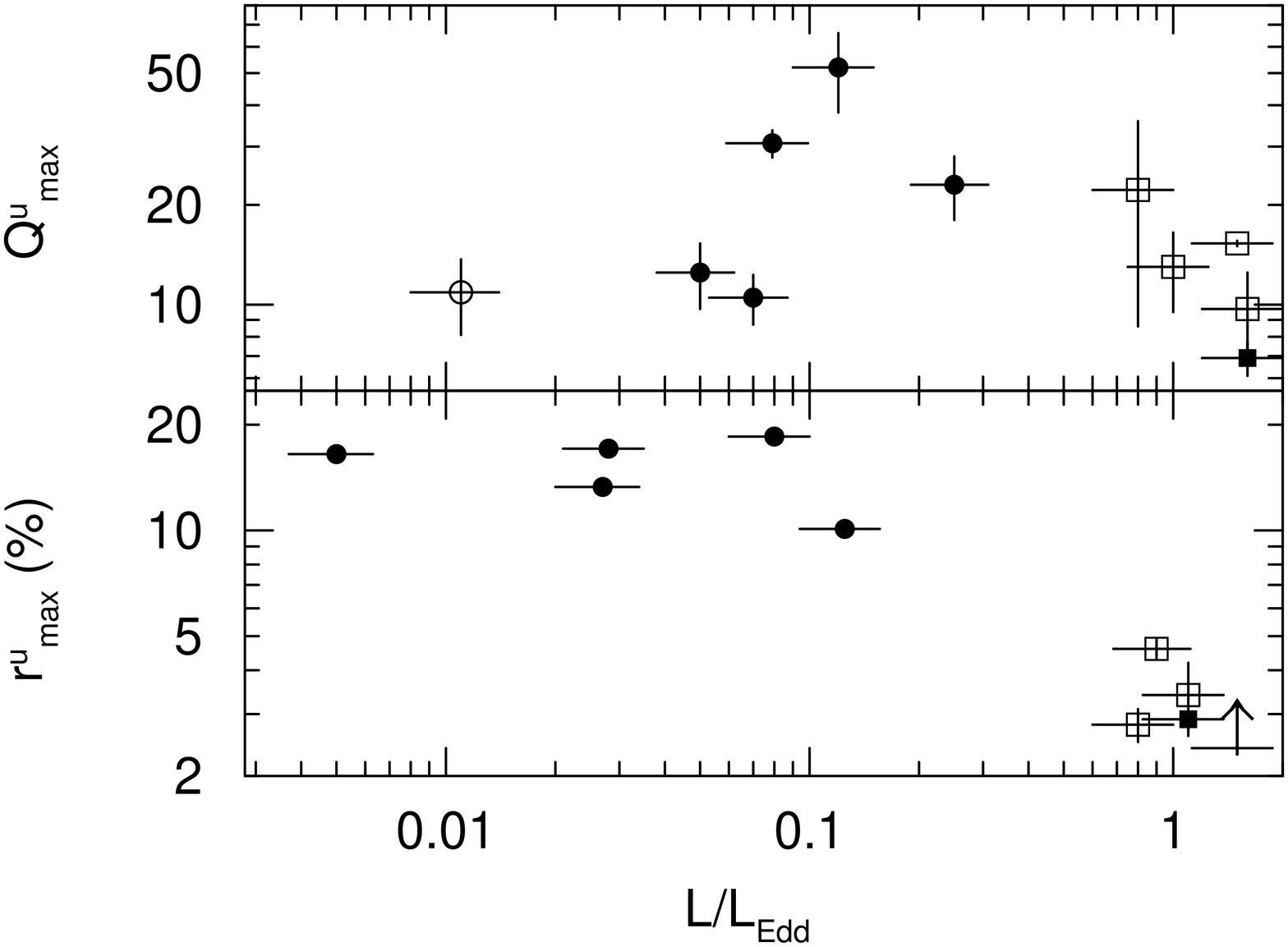,angle=-90,width=7.0cm}}
\caption{{\em Left panel}: The maximum coherence (upper panel) and
maximum rms amplitude (lower panel) of the lower kHz QPO as a function
of the source luminosity at the time at which those maximum values were
reached. The luminosity is in  units of the Eddington luminosity for a
$1.9 \msun$ neutron star. Filled symbols indicate measurements over the
full energy band covered by the PCA on board {\em RXTE}. Open circles
indicate measurements over a limited energy band; in these cases (except
for Sco X-1), the rms amplitudes have been divided by 1.25
\citep[see][for details]{mendez-06}. The rms amplitude in the case of
Sco X-1 is not corrected for dead-time, and hence is only a lower limit
(indicated with an arrow pointing upwards in the lower panel). Circles
are Atoll sources; squares are Z sources. {\em Right panel}: The maximum
coherence (upper panel) and maximum rms amplitude (lower panel) of the
upper kHz QPO as a function of the source luminosity at the time at
which those maximum values were reached. The luminosity is in units of
the Eddington luminosity for a $1.9 \msun$ neutron star. Symbols are
the  same as in the left panel. \label{low}}
\end{figure*}


\section{Results}
\label{results}


Figure~\ref{low} shows the dependence of the maximum coherence and
maximum rms amplitude of both kHz QPOs as a function of source
luminosity (in units of the Eddington luminosity for a $1.9 \msun$
neutron star). From this Figure it is apparent that $r^{\rm u}_{\rm
max}$ and $r^{\ell}_{\rm max}$ both decrease more or less exponentially
with $L/L_{\rm Edd}$. Using roughly the same sample of sources,
\cite{jonker-0918} had already noticed that the rms amplitude of the
upper kHz QPO decreases as the luminosity of the source increases.


\begin{figure*}
\centerline{\epsfig{file=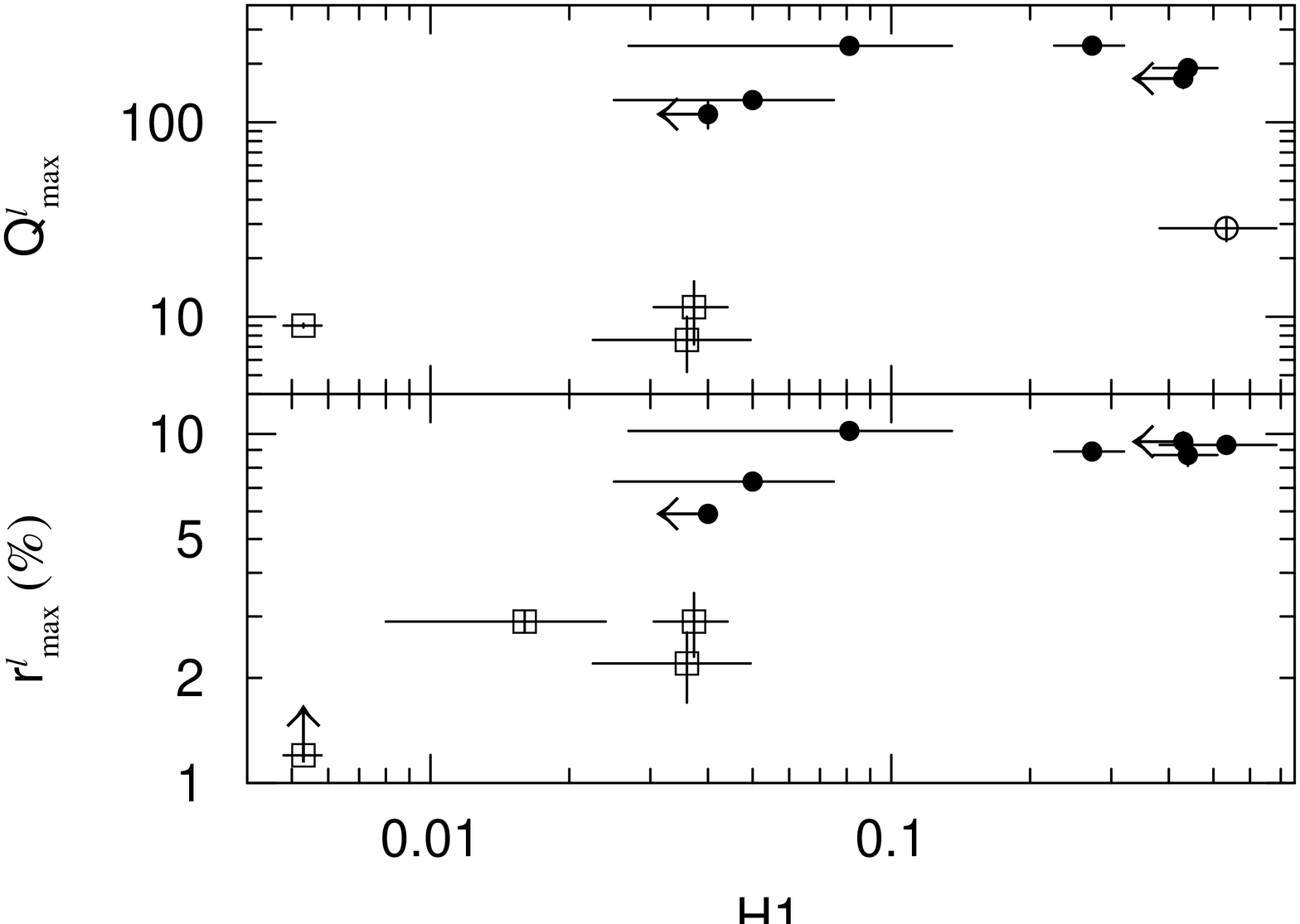,angle=-90,width=7.0cm}~~~~~~
            \epsfig{file=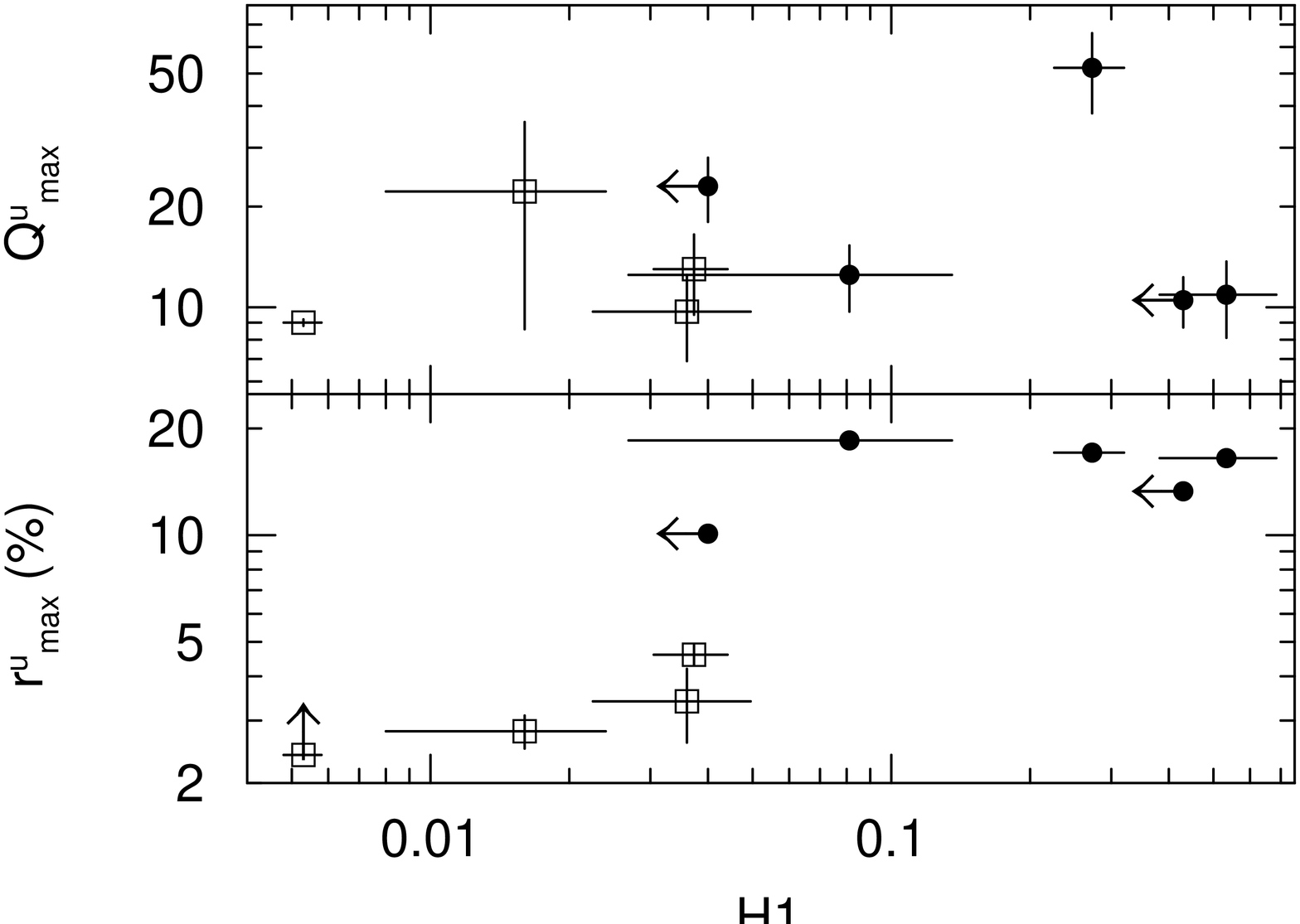,angle=-90,width=7.0cm}}
\caption{The maximum coherence (upper panels) and maximum rms amplitude
(lower panels) of the lower kHz QPO (left) and the upper kHz QPO (right)
as a function of the average source hardness, H1, defined as the ratio
of the $40-80$ keV to the $13-25$ keV count rate, measured with {\em
HEAO-1} \citep{levine-heao-1, vanparadijs-vanderklis}. For an
explanation of the symbols  see the caption of Figure~\ref{low}. The rms
amplitudes in the case of Sco X-1 are not corrected for dead-time, and
hence are lower limits (indicated with a vertical arrow in the lower
panels). Upper limits to the hardness are indicated with horizontal
arrows pointing to the left. I do not include GX 5-1 in this Figure
because the {\em HEAO-1} measurements of this source suffer
contamination from a previously unknown hard X-ray source in the field
\citep[see explanation in][]{levine-heao-1}.\label{h1}} 
\end{figure*}

From the same Figure, it is also apparent that at low luminosity,
$Q^{\ell}_{\rm max}$ first increases with $L$ up to $L/L_{\rm Edd} \sim
0.04$, and then it decreases exponentially. On the other hand, $Q^{\rm
u}_{\rm max}$ does not show any significant trend with luminosity.

The gap in Figure~\ref{low} at $L/L_{\rm Edd} \sim 0.25 - 0.7$ separates
the Atoll sources at low $L$, and the Z sources at high $L$
\cite[see][for a definition of Atoll and Z sources]{hk89}. That gap
would be occupied by the four intermediate-type sources, GX 9+1, GX 9+9,
GX 3+1, and GX 13+1, which so far have not shown any kHz QPOs
\citep*{strohmayer-gx-sources, wijnands-gx-sources, homan-13+1,
oosterbroek-3+1}. The upper limit to the rms amplitude of the QPO in
these sources ranges from 1.6\% to 2.6\%. Since the range of
luminosities spanned by these sources \citep{christian-swank} is
$L/L_{\rm Edd} \sim 0.12 - 0.44$, these upper limits are much lower than
would be expected from the interpolation of the trends of $r^{\ell}_{\rm
max}$ and $r^{\rm u}_{\rm max}$ with $L/L_{\rm Edd}$ in Figure
\ref{low}.

From Figure~\ref{low} it is apparent that the maximum rms amplitude of
both kHz QPOs and the maximum coherence of the lower kHz QPO are
consistently lower in the Z sources than in the Atoll sources (see also
below). Since the detection of QPOs in Z sources generally require
calculating average power spectra over longer time intervals than for
Atoll sources, this difference could in principle be due to the
frequency drift of the QPOs, which would artificially reduce their $Q$
values. However, measurements of the coherence of the lower kHz QPO in
the Atoll source 4U 1608--52 and the Z source Sco X-1 on short time
intervals (less than a few hundred seconds), when $\nu_{\ell} \sim 600$
Hz, yields $Q_{\ell} = 74.0 \pm 4.6$ for 4U 1608--52, and $Q_{\ell} =
4.2 \pm 0.4$ for Sco X-1. This shows that in the Z source Sco X-1, even
over very short time intervals, the lower kHz QPO is intrinsically
significantly broader than in the Atoll source 4U 1608--52. 

From Figure \ref{low} \citep[see also][]{mendez-06} it is also apparent
that, except for the case of 4U 0614+09, $Q^{\ell}_{\rm max}$ and
$r^{\ell}_{\rm max}$ are positively correlated with each other.
Concerning 4U 0614+09, the hardest source in the sample and the one at
the lowest luminosity, it is as if in this case $Q^{\ell}_{\rm max}$
were too low for $r^{\ell}_{\rm max}$, or conversely, as if
$r^{\ell}_{\rm max}$ were too high for $Q^{\ell}_{\rm max}$. Recent work
\citep{barret-4srcs} confirms that the decrease of $Q^{\ell}_{\rm max}$
at low luminosity is real. 


Van Paradijs and van der Klis \cite*{vanparadijs-vanderklis} have shown
that there is a general correlation between the average source
luminosity and the average source hardness, H1, defined as the ratio of
the count rate in the $40-80$ keV band to the count rate in the $13-25$
keV band, measured with {\em HEAO-1} \citep[see][]{levine-heao-1}. In
that respect, as expected, the plots of $Q_{\rm max}$ and $r_{\rm max}$
vs. hardness in Figure~\ref{h1} show that $r^{\rm u}_{\rm max}$ and
$r^{\ell}_{\rm max}$ both increase with the hardness ratio H1,
$Q^{\ell}_{\rm max}$ increases with H1 and then it decreases for 4U
0614+09, the hardest source in this sample, and as in the plots as a
function of luminosity, $Q^{\rm u}_{\rm max}$ is consistent with being
constant with H1. I caution the reader that contrary to the $L/L_{\rm
Edd}$ values that I plot in Figure~\ref{low}, for this Figure I use
average values of the spectral hardness measured several years before
the kHz QPOs were discovered. Notice that GX 340+0, for which the
distance, and hence the luminosity, is uncertain, appears in
Figure~\ref{h1} close to the other Z sources (open symbols; GX 340+0 is
the point at H1 $= 0.036$, just to the left of GX 17+2 at H1 $= 0.037$);
since H1 is a distance-independent parameter, this suggests that the
distance to GX 340+0 is not too much in error.

It is interesting to notice in this Figure that there is no gap in the
distribution of sources as a function of hardness between the Z and
Atoll sources. This is opposite to what is apparent in the plot of QPO
parameters vs. luminosity, where there is a gap corresponding to the
intermediate-type, the GX-sources (see above). The H1 values of the
GX-sources range from 0.03 to 0.14. 

Both in Figure \ref{low} and \ref{h1} there appears to be a gap in the
values of $Q^{\ell}_{\rm max}$, and perhaps also $r^{\ell}_{\rm max}$
and $r^{\rm u}_{\rm max}$ between the Z and Atoll sources. I have just
shown that this difference is real, and is not an artifact due to the
way the QPOs are measured. While this difference may indicate a
dependence on luminosity (see Fig.~\ref{low}) or on spectral hardness
(see Fig.~\ref{h1}), this could also point to a difference between Z and
atoll sources. It is worth noting, however, that there is still a trend
of $Q^\ell_{\rm max}$ and both rms amplitudes {\em within} the atoll
sources in Figure~\ref{low}. Furthermore, there is a significant trend
in the relations between $Q_{\rm max}$ and $r_{\rm max}$ within the
atoll sources; e.g., the relation between $Q^\ell_{\rm max}$ and
$r^\ell_{\rm max}$ is $8\sigma$ different from a constant
\citep[see][for details]{mendez-06}. All this suggests that the
distinction between Z and Atoll sources cannot be the (only) explanation
for this difference.


\section{Discussion}
\label{discussion}


I study the maximum amplitude, $r_{\rm max}$, and maximum coherence,
$Q_{\rm max}$, of the kHz QPOs as a function of luminosity and hardness
for a large sample of low-mass X-ray binaries. I show that the maximum
coherence of the lower kHz QPO, $Q^{\ell}_{\rm max}$, first increases up
to $L \sim 0.04 L_{\rm Edd}$ and then decreases with luminosity, whereas
the maximum coherence of the upper kHz QPO, $Q^{\rm u}_{\rm max}$, is
independent of luminosity. I also find that the maximum rms amplitudes
of both the lower and the upper kHz QPOs, $r^{\ell}_{\rm max}$ and
$r^{\rm u}_{\rm max}$, respectively, decrease monotonically with
luminosity and increase monotonically with the hardness of the source.
\citep[The dependence of $r^{\rm u}_{\rm max}$ on luminosity and
hardness was first reported by][]{jonker-0918}.


From the above results it follows that for all sources, $r^{\ell}_{\rm
max}$ and $r^{\rm u}_{\rm max}$ are positively correlated with each
other. Also, for all sources with $L \simmore 0.04 L_{\rm Edd}$, that is
all sources in this paper except 4U 0614+09, the hardest source in the
sample and the one at the lowest luminosity, $Q^{\ell}_{\rm max}$ is
positively correlated both with $r^{\ell}_{\rm max}$ and $r^{\rm u}_{\rm
max}$. $Q^{\rm u}_{\rm max}$ is independent of $Q^{\ell}_{\rm max}$ or
the maximum rms amplitude of the kHz QPOs.


In individual sources, both $r_{\ell}$ and $Q_{\ell}$ increase with
$\nu_{\ell}$ and then drop rather abruptly at the high end of the
$\nu_{\ell}$ range; $r_{\rm u}$ also increases and then drops at high
$\nu_{\rm u}$ values, and $Q_{\rm u}$ is more or less constant or
increases slightly with $\nu_{\rm u}$ \cite[e.g.,][]{disalvo-1728,
disalvo-1636, mendez-3srcs}. In the case of 4U 1636--53, \cite{barret-1636}
interpret the sudden drop of the coherence and rms amplitude of
$L_{\ell}$, together with the existence of a frequency above which
$L_{\ell}$ is not detected, as evidence of the innermost stable circular
orbit around the neutron star in this system.


From the results of individual sources and those of the sample of
sources that I present in this paper, it is apparent that the behaviour
of the coherence and rms amplitude of the kHz QPOs as a function of the
{\em QPO frequency} in {\em individual sources} is similar to the
behaviour of the maximum coherence and maximum rms amplitude of the kHz
QPOs as a function of {\em luminosity} in the {\em sample of sources}.
Since in individual sources there is a general relation between QPO
frequencies and source intensity, in the sense that at higher intensity
the QPOs generally appear at higher frequencies (but remember the
parallel-track effect), this raises the question of whether the same
mechanism may be behind both behaviours.

The link between these two behaviours need not be luminosity, but could
be the high-energy emission (or hardness) in these systems. On one hand,
in the sample of sources the maximum coherence and maximum rms amplitude
of the kHz QPO, except the maximum rms amplitude of $L_{\rm u}$, appear
to correlate fairly well with spectral hardness (see Figure \ref{h1}),
while on the other hand in individual sources the QPO frequencies are
well correlated with the spectral hardness of the source
\citep{mendez-1608}, the index of the power law that fits the
high-energy part of the X-ray spectrum \citep{kaaret-0614-1608}, or $S$,
a parameter that measures the position along the track traced out by the
source in a colour-colour or colour-intensity diagram \citep[called
$S_z$ and $S_a$ in the Z and Atoll sources, respectively; see
e.g.,][]{jonker-340+0, mendez-1728}. From this, it follows that in
individual sources there should be a relation between QPO rms amplitude
and coherence on one hand and spectral hardness on the other.


The comparison between individual sources and the sample of sources
suggests that the same mechanism is responsible for the drop of
coherence and rms amplitude of the lower kHz QPO with QPO frequency in
individual sources as well as for the drop of maximum QPO coherence and
maximum QPO rms amplitude with luminosity in the sample of sources. Most
likely the mechanism is related to the high-energy emission in these
systems. This does not necessarily mean that the fractional emission at
high energies (represented by the hardness or X-ray colours) is the root
mechanism that drives all QPO parameters (QPO frequency, coherence, and
rms amplitude). For instance, one possibility (there could be many
others) is that the (instantaneous) mass accretion rate sets the size of
the inner radius of the disc \citep{vanderklis-2001}, which in turn
determines the QPO frequency as well as the relative contribution of the
high-energy part of the spectrum to the total luminosity. If the
efficiency of the modulation mechanism (related to the rms amplitude
$r$) and the lifetime of the oscillations (related to the coherence $Q$)
that produce the QPO depended upon the emission from the high-energy
part of the source spectrum \citep[see][for a discussion of possible
ways in which this could happen]{mendez-06}, observationally it would
appear as if the coherence and rms amplitude of the QPO depended upon
the QPO frequency, and hence upon the radius in the disc at which the
QPO is produced. The sudden drop of the coherence and rms amplitude of
the QPO at some QPO frequency would then appear to be associated to a
dynamical peculiarity in the accretion disc, for instance the ISCO.
Observing the same source repeatedly would not allow to distinguish the
above scenario from one in which QPO coherence and rms amplitude were
actually set by QPO frequency or the dynamics in the accretion disc.

To distinguish between these two options, one would need to observe a
sample of sources of kHz QPOs for which the mass-accretion rate, and
hence the relative contribution of the high-energy part of the spectrum
to the total emission, was different from source to source. In that
case, QPO coherence and rms amplitude would drop for sources accreting
mass at higher rates, even if the frequency of the QPO was more or less
the same from source to source. Since, as I have shown in this paper,
this is the general behaviour observed in sources of kHz QPOs, it is
reasonable to infer that a mechanism similar to the one I described in
the previous paragraph is effective in setting the coherence and rms
amplitude of the kHz QPOs. If this is correct, this also implies that
the drop of QPO coherence and rms amplitude as a function of QPO
frequency in individual sources cannot be due to effects of the ISCO.


Note also that in individual sources not just the rms amplitude of the
kHz QPOs, but also the rms amplitude of other lower-frequency QPOs
decrease with increasing QPO frequency. For instance, in four Atoll
sources, 4U 1728--34, 4U 1608--52, 4U 0614+09, and 4U 1636--53, the rms
amplitudes of the ``bump'', a QPO at $\sim 0.1-30$ Hz, the ``hump'', a
QPO at $\sim 1-40$, and the hectohertz QPO at $\sim 100-300$ Hz, all
drop as the frequencies of the kHz QPOs increase, in a similar fashion
as the amplitude of the upper and lower kHz QPOs \citep[e.g.,][and
references therein; see also there a description of these other
QPOs]{altamirano-1636}. This also argues against the interpretation of
the ISCO as the cause of the drop of the rms of the kHz QPOs, and
indicates that the amplitudes of {\em all} variability components are
set by the same mechanism which, as I suggested, could be the same one
that governs the high-energy spectral component.


\section{Conclusion}
\label{conclusion}

I study the maximum amplitude, $r_{\rm max}$, and maximum coherence,
$Q_{\rm max}$, of the kHz QPOs as a function of luminosity and hardness
for a dozen low-mass X-ray binaries. I find that: (i) The maximum
coherence of the lower kHz QPO, $Q^{\ell}_{\rm max}$, first increases up
to $L \sim 0.04 L_{\rm Edd}$ and then decreases with luminosity. (ii)
The maximum coherence of the upper kHz QPO, $Q^{\rm u}_{\rm max}$, is
independent of luminosity. (iii) The maximum rms amplitudes of both the
lower and the upper kHz QPOs, $r^{\ell}_{\rm max}$ and $r^{\rm u}_{\rm
max}$, respectively, decrease monotonically with luminosity. (iv) Both
$r^{\rm u}_{\rm max}$ and $r^{\ell}_{\rm max}$ increase with the source
hardness, $Q^{\ell}_{\rm max}$ first increases with hardness and then
drops for the hardest source in the sample, and $Q^{\rm u}_{\rm max}$ is
independent of hardness. (v) The relation between $Q_{\rm max}$ and
$r_{\rm max}$ with luminosity in the sample of sources is similar to the
relation between $Q$ and $r$ with QPO frequency in individual sources.
(vi) The above argues against the interpretation that the drop of QPO
coherence and QPO rms amplitude at high QPO frequency in individual
sources is due to effects related to the innermost stable orbit around
the neutron star in these systems.


\section*{Acknowledgments}

I thank Didier Barret and Cole Miller for valuable discussions on the
ideas that I present in this paper. I am grateful to Manuel M\'endez for
his collaboration, and to Ahmed and Lidy Helmi for their hospitality
during the period in which I did this work. This research has made use
of data obtained through the High Energy Astrophysics Science Archive
Research Center Online Service, provided by the NASA/Goddard Space
Flight Center. The Netherlands Institute for Space Research (SRON) is
supported financially by NWO, the Netherlands Organisation for
Scientific Research.


\end{document}